# Threshold voltage instability by charge trapping effects in the gate region of p-GaN HEMTs

Giuseppe Greco[1*], Patrick Fiorenza[1*], Filippo Giannazzo[1], Corrado Bongiorno[1], Maurizio Moschetti[2], Cettina Bottari[2], Mario Santi Alessandrino[2], Ferdinando Iucolano[2] and Fabrizio Roccaforte[1]

[1]*Consiglio Nazionale delle Ricerche – Istituto per la Microelettronica e Microsistemi, Strada VIII, n. 5 - Zona Industriale, 95121 Catania, Italy*
[2]*STMicroelectronics, Stradale Primosole 50, 95121 Catania, Italy*

Corresponding author: giuseppe.greco@imm.cnr.it

**ABSTRACT**

In this work, the threshold voltage instability of normally-off p-GaN high electron mobility transistors (HEMTs) has been investigated by monitoring the gate current density during device on-state. The origin of the gate current variations under stress has been ascribed to charge trapping occurring at the different interfaces in the metal/p-GaN/AlGaN/GaN system. In particular, depending on the stress bias level, electrons ($V_G < 6$ V) or holes ($V_G > 6$ V) are trapped, causing a positive or negative threshold voltage shift $\Delta V_{TH}$, respectively. By monitoring the gate current variations at different temperatures, the activation energies associated to the electrons and holes trapping could be determined and correlated with the presence of nitrogen (electron traps) or gallium (hole traps) vacancies. Moreover, the electrical measurements suggested the generation of a new electron-trap upon long-time bias stress, associated to the creation of crystallographic dislocation-like defects extending across the different interfaces (p-GaN/AlGaN/GaN) of the gate stack.

Nitride based heterostructures (e.g. AlGaN/GaN, InAlGaN/GaN, AlN/GaN, etc.) are considered key materials for the next generation of high-power and high-frequency electronics [1]. Specifically, for high-voltage applications, the wide bandgap, large critical electric field and saturation velocity of these nitride materials are important features, enabling to reduce the on-state resistance ($R_{ON}$) and increase the breakdown voltage ($V_B$). Moreover, the presence of the two-dimensional electron gas (2DEG), formed e.g. at AlGaN/GaN interfaces, offers the possibility to fabricate high electron mobility transistors (HEMTs) with a high channel mobility, operating at high frequency.

Due to the spontaneous formation of the 2DEG, HEMTs are inherently normally-on devices, and a negative gate voltage has to be applied in order to turn-off the device. However, in order to guarantee a safe operation in power conversion systems, normally-off devices are highly desired [2].

While several approaches to fabricate normally-off GaN-based HEMTs have been proposed in the literature [3,4,5,6], the p-GaN gate HEMT is currently the only one available on the market. In such a device, when a p-GaN layer is put in contact with the AlGaN barrier layer, a net transfer of electrons occurs from the AlGaN/GaN heterostructures to the p-GaN, which results in the formation of a space charge region between the p-GaN and the AlGaN. At the equilibrium, for appropriate thickness and Al concentration of the AlGaN layer [7], the Fermi level at the AlGaN/GaN interface will lie below the conduction band edge of the AlGaN, thus leading to the complete depletion of the 2DEG. Hence, a positive threshold voltage $V_{TH}$ will have to be applied to the p-GaN gate of the HEMT to restore the charge in the 2DEG channel. Despite normally-off HEMTs with p-GaN gate are already commercially available, their performances are often affected by reliability concerns, like a high gate current level in the on-state, which can cause an early degradation, with a consequent reduction of the device lifetime [8,9]. In such devices, the mechanisms of the gate current leakage have been widely investigated. Depending on the applied gate bias, Space Charge Limited Current [10], Trap Assisted Tunnelling [11] or Thermionic Field Emission model [8] have been identified as dominant mechanisms for gate current conduction.

Gate current transient measurements, correlated with time-to-failure (TTF) analysis are typically used to predict the device lifetime [11,12,13,14,15]. In this case, accelerated device failure tests at gate bias levels higher ($V_G$ > 8 V) than the device operational gate bias (6 V) are typically used [16,17]. At these bias levels, gate current transient measurements can give useful information on the $V_{TH}$ [8,10,16,17]. However, the presence of several interfaces in the metal/p-GaN/AlGaN/GaN system often makes it difficult to identify the specific trapping phenomena responsible for the $V_{TH}$ instability.

*Tang et al.* [16] attributed the positive $V_{TH}$ shift in p-GaN HEMTs to electron trapping occurring at the p-GaN/AlGaN interface. On the other hand, *He et al.* [8] explained these effects by a hole depletion in the p-GaN region, or equivalently, by electron trapping in the p-GaN depletion region. Recently, *Shi et al.* [10] used very fast gate current transient analysis ($10^{-4}$-$10^{0}$ s) to monitor $V_{TH}$ shift during fast device switch-on. From the gate current variation after stress, they evaluated the residual net trapped charges in the gate-stack, concluding that either holes or electrons can be trapped in the gate region, depending on the stress gate bias [10]. However, the origin of holes or electrons trapping phenomena was not well understood and no information on the energy trap levels was provided.

In this context, studying the $V_{TH}$ instability under bias stress values far from the device breakdown condition may help to discern between trapping phenomena of different origin. Moreover, long stress current transients could give quantitative information on the charge trapping mechanisms and on their influence on $V_{TH}$ instability.

In this letter, the $V_{TH}$ stability of p-GaN HEMTs has been investigated by gate current transient acquired at different $V_G$ bias stress, from low- to high-bias condition. The gate current variations as a function of the stress-time allowed the comprehension of the charge trapping occurring at the gate regions. Indeed, depending on the applied gate bias, electron or hole trapping process can play a key role for the reliability of the p-GaN HEMTs.

Normally-off p-GaN/AlGaN/GaN HEMTs on Si substrates were used in this work. The gate structure consisted of a 18-nm-thick AlGaN barrier layer with a 20% Al content, and a 90 nm thick p-GaN layer with a nominal Mg concentration of $3\times10^{19}$ $cm^{-3}$. From Capacitance-Voltage measurements [18] an acceptor concentration of $5\times10^{18}$ $cm^{-3}$ has been estimated. Then, by assuming an ionization energy for Mg-dopant of about 160 meV [19], typical hole concentration values in the order of $10^{17}$ $cm^{-3}$ are expected at room temperature. In our device, a TiN-based Schottky gate contact was used, while Ta-based electrodes have been formed for source-drain Ohmic contacts [20]. The device electrical characterization has been carried out in a Karl–Suss MicroTec probe station equipped with a parameter analyser. Scanning transmission electron microscopy (STEM) analyses on cross sectioned p-GaN HEMT devices structures were performed with a JEOL ARM200C at 200 keV.

Firstly, the $V_{TH}$ stability of p-GaN HEMTs has been investigated. For this purpose, the transfer characteristics of the p-GaN HEMTs have been acquired before and after bias stress test. The stress test consisted in applying a positive gate bias ($V_G$), in the range between +2 V and +9.5 V, for 4000 s. For each condition, a "fresh" p-GaN HEMT (not stressed previously) was used. Fig. 1(a) and Fig. 1(b) report the transfer characteristics of the p-GaN HEMTs before and after gate bias stress ($t_{stress}$ = 4000 s) carried out at lower (2 V < $V_G$ < 6 V) and higher gate bias (6 V < $V_G$ < 9.5 V), respectively. At $V_G$ = 9.5 V the stress has been interrupted after 1920 s, due to a rapid increases of the gate current resembling a "soft-breakdown" event. Fig. 1(c) shows the threshold voltage shift $\Delta V_{TH}$ as function of the applied gate bias stress, defining $V_{TH}$ as the gate bias corresponding to a drain current of 1 µA. As can be seen, in all cases after stress measurement, a positive shift of $V_{TH}$ has been detected. However, $\Delta V_{TH}$ increases by increasing $V_G$ stress from 2 V to 6 V, and it starts to decreases at higher $V_G$.

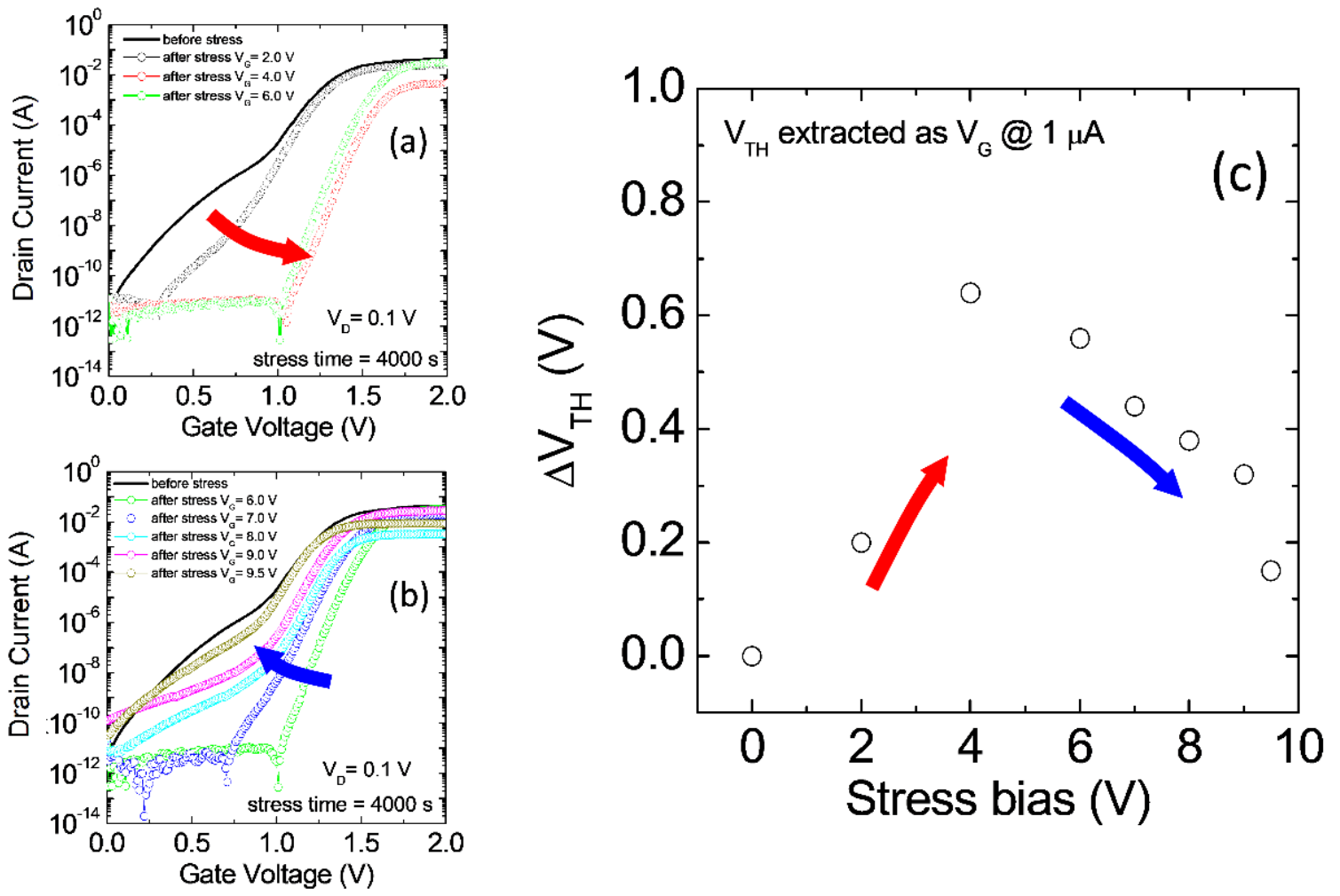

Fig. 1: Transfer characteristics $I_D$-$V_G$ of the p-GaN HEMTs before and after low (a) and high (b) gate bias stress $V_G$. Threshold voltage variation $\Delta V_{TH}$ as a function of the gate bias stress (c). The duration of the gate bias stress was $t_{stress}$=4000s.

The gate current density ($J_G$) is reported in Fig. 2(a) and Fig. 2(b) as a function of the stress time, for low (2 - 6 V) and high (7 - 9.5 V) values of the positive gate bias stress ($V_G$), respectively. As expected, for higher applied $V_G$ a higher gate current density is detected. For the highest bias stress of $V_G$ = 9.5 V, the device breakdown is well visible, since it occurs after 1920 s, i.e. well before the end of the 4000 s stress measurement duration.

To better highlight the change of the current density with time, the percentage variation of $J_G$ with respect to its initial value, i.e. $[J_{G(t)}-J_{G(0)}]/J_{G(0)}$, it is displayed in Fig. 2(c). Here, one can easily observe that $J_G$ decreases with the time for low gate bias values (2 V < $V_G$ < 6 V), while it increases at high gate bias values, ($V_G$ > 6 V). Furthermore, for the highest bias stress values of $V_G$ = 9 V and $V_G$=9.5 V, the gate current $I_G$ increases for a certain time and then starts decreasing, reaching eventually a "soft-breakdown" event at $V_G$=9.5 V. The changes observed in gate current density can be correlated with electrons or holes trapping in the gate region. Indeed, when an electron is captured in the gate stack, the increased electrostatic potential will lead to a higher barrier for electrons. Hence, a higher gate voltage has to be applied to reach the same gate current level. Conversely, if holes are captured, the electrostatic potential decreases and a lower gate bias will be required to obtain the same gate current. In fact, monitoring the gate current variation is a powerful method to monitor the electron trapping in the gate stack [21, 22, 23]. In our case, for gate bias stress between 2 V and 6 V, a decrease of the current density is observed, which can be correlated with an electron trapping. Instead, for gate bias stress higher than 6 V, an increase of the $J_G$ is observed, which is caused by a holes trapping. Hence, the behaviour observed in the $J_G$ transient acquired at $V_G$ = 9 V and $V_G$=9.5 V could be explained by a change in the total net trapping charges that from positive (holes) become negative (electrons) after a certain time. The origin of this variation can be justified by the generation of new electron trapping states, which exceed those observed for holes. Moreover, in case of stress at low applied $V_G$, the recovery of the $V_{TH}$ close to its original value can be stimulated by applying a negative $V_G$ stress or a suitable thermal budget [24,25]. This possibility is generated by the reversible nature of the trapping and detrapping phenomena. Different is the case at high $V_G$ bias stress (as $V_G$ = +9.5 V), where new traps have been generated, inducing a degradation on the stressed device. Indeed, the generation of new traps not only influences the device $V_{TH}$ stability but also affects the mechanism involved in device breakdown and the device lifetime

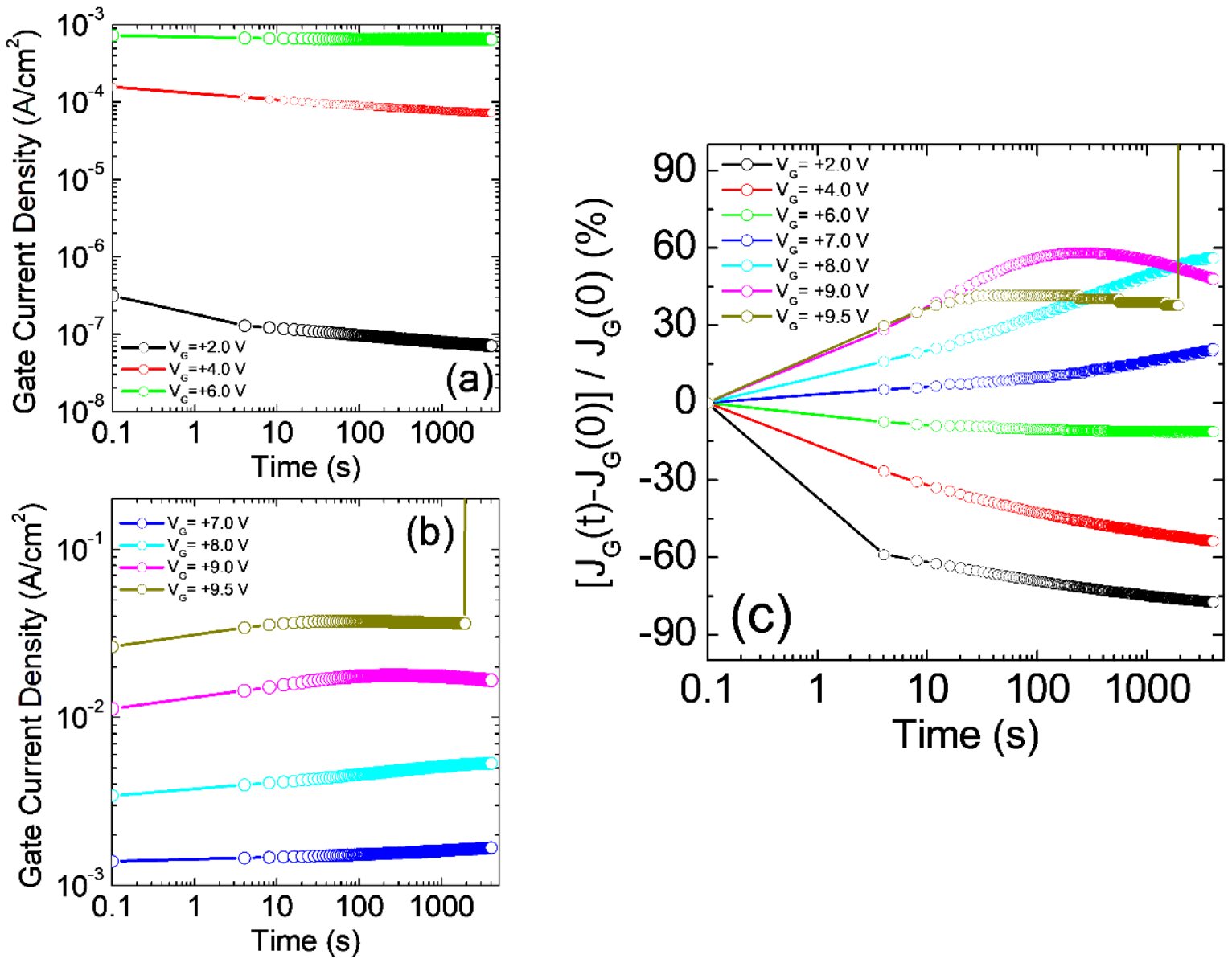

Fig. 2: Gate current density transient ($J_G$-t) collected by applying a low gate bias stress ($V_G$= 2-6 V) (a) and a high gate bias stress ($V_G$= 7-9.5 V) (b); Percentage variation of $J_G$ (defined as $[J_{G(t)}-J_{G(0)}]/J_{G(0)}$) as function of the measurement stress time at the different applied gate bias (c).

In order to acquire information on the origin of the observed trapping phenomena, we examined in more detail the transient measurements acquired at $V_G$=+2 V (low stress bias) and $V_G$=+8 V (high stress bias).

According to the model proposed by *Joh and Del Alamo* [26], the time dependent gate current I(t) can be described by:

$$I(t) = I_\infty - \sum_i A_i \cdot exp\left(-\frac{t}{\tau_{trap_i}}\right) \qquad \text{Eq. 1}$$

where $I_\infty$ is the current estimated at t= ∞, $A_i$ and $\tau_i$ are the amplitude and the time constant of the detected trapping process. The sign of the amplitude enables to distinguish between electron ($A_i < 0$) or hole ($A_i > 0$) trapping, while the time constant $\tau_i$ is a parameter depending on the trap electrical properties as well on its energy position in the bandgap. Fig. 3(a) and Fig. 3(b) show the gate current density as function of time for $V_G$= +2 V and $V_G$= +8 V, respectively. From the fits of the experimental data carried out using Eq. 1, the time constants of the different trapping process (electron or hole) have been estimated.
To acquire information on the energy trap level $\Delta E_{trap}$ of these processes, the same analysis has been performed at different temperatures between 25 °C and 100 °C. Indeed, the time constant $\tau_{trap}$ characteristic of a trap level $\Delta E_{trap}$ depends on the temperature by the equation

$$\tau_{trap} = \frac{1}{\sigma_{trap} v_{th} N_C exp\left(\frac{\Delta E_{trap}}{kT}\right)} \qquad \text{Eq. 2}$$

with $\sigma_{trap}$ the electron capture cross section of the trap level, $v_{th}$ the electron thermal velocity and $N_C$ the effective density of states at the conduction band

$$N_C = 2\left(\frac{2\pi m^* kT}{h^2}\right)^{\frac{3}{2}} \qquad \text{Eq. 3}$$

Combining Eq.2 and Eq.3 and considering the quadratic dependence of $v_{th}$ on the temperature, it is possible determine the following Arrhenius-like dependence for $\tau_{trap}$:

$$\tau_{trap} T^2 = \frac{1}{C} \cdot exp\left(-\frac{\Delta E_{trap}}{kT}\right) \qquad \text{Eq. 4}$$

with $C = \frac{1}{2\sqrt{3}\left(\frac{2\pi}{h^2}\right)^{\frac{3}{2}} k^2 m^* \sigma_{trap}}$.

The inset of Fig. 3(a) and Fig. 3(b) show the gate current transient measurement at $V_G$= +2 V and $V_G$= +8 V acquired in the temperature range between 25 °C and 100 °C. From the fit of these measurements the time constants related to the gate stress bias of $V_G$=+2V and $V_G$= +8 V have been calculated. Then, the Arrhenius plots, calculated with Eq. 4 in the case of $V_G$= +2 V and $V_G$= +8 V are displayed in Fig. 3(c) and Fig. 3(d), respectively. For $V_G$= +2 V, two different traps levels for electron have been estimated, i.e. at the energy levels of 67 meV and 83 meV from the conduction band. In the case of $V_G$= +8 V from the Arrhenius plot an energy level for hole trap of 410 meV from the valence band has been calculated. While the trapping phenomena and their activation energies are typically associated to the presence of point defects in the material, the electrical measurements only cannot reveal their physical nature. However, several literature works already described the trapping phenomena in AlGaN/GaN heterostructures by combining first principles calculations with optical analyses [27,28,29,30,31]. Basing on this wide literature database, the energy levels estimated from our gate current transient analysis are plausibly correlated with the presence of nitrogen and gallium vacancies. In fact, the presence of electron traps located in the energy level around 70-80 meV below the conduction band, can be attributed to the presence of nitrogen vacancy in the GaN or AlGaN layer [32]. On the other hand, the deep trap states for holes, estimated around 400 meV from the valence band, have been observed in literature in presence of gallium vacancies [33,34].

To elucidate the electron and holes trapping effects it is important to distinguish between the different gate conduction mechanisms acting during the on-state of the device. Indeed, the gate region of the p-GaN stack comprises two different interfaces, i.e. metal/p-GaN and p-GaN/AlGaN interface. For positive applied gate bias, the p-GaN/AlGaN junction is forward biased while the metal/p-GaN junction is reversed biased. Then, at lower $V_G$, electrons are driven from the 2DEG-channel towards the p-GaN gate, by the emission through the AlGaN barrier layer. A fraction of these electrons will reach the p-GaN recombining with the holes of this p-type layer, while a remaining part will be captured by the traps in the AlGaN barrier layer. This scenario is schematically depicted in the inset of Fig. 3(c). On the other hand, by increasing $V_G$, most of the bias is applied on the metal/p-GaN interface, which is reversed biased. Here, holes are injected from the metal/p-GaN interface towards the p-GaN/AlGaN interface. In this scenario, part of the holes are trapped into the p-GaN layer, causing the increase of the $J_G$ observed during stress measurements (see inset of Fig. 3(d)).

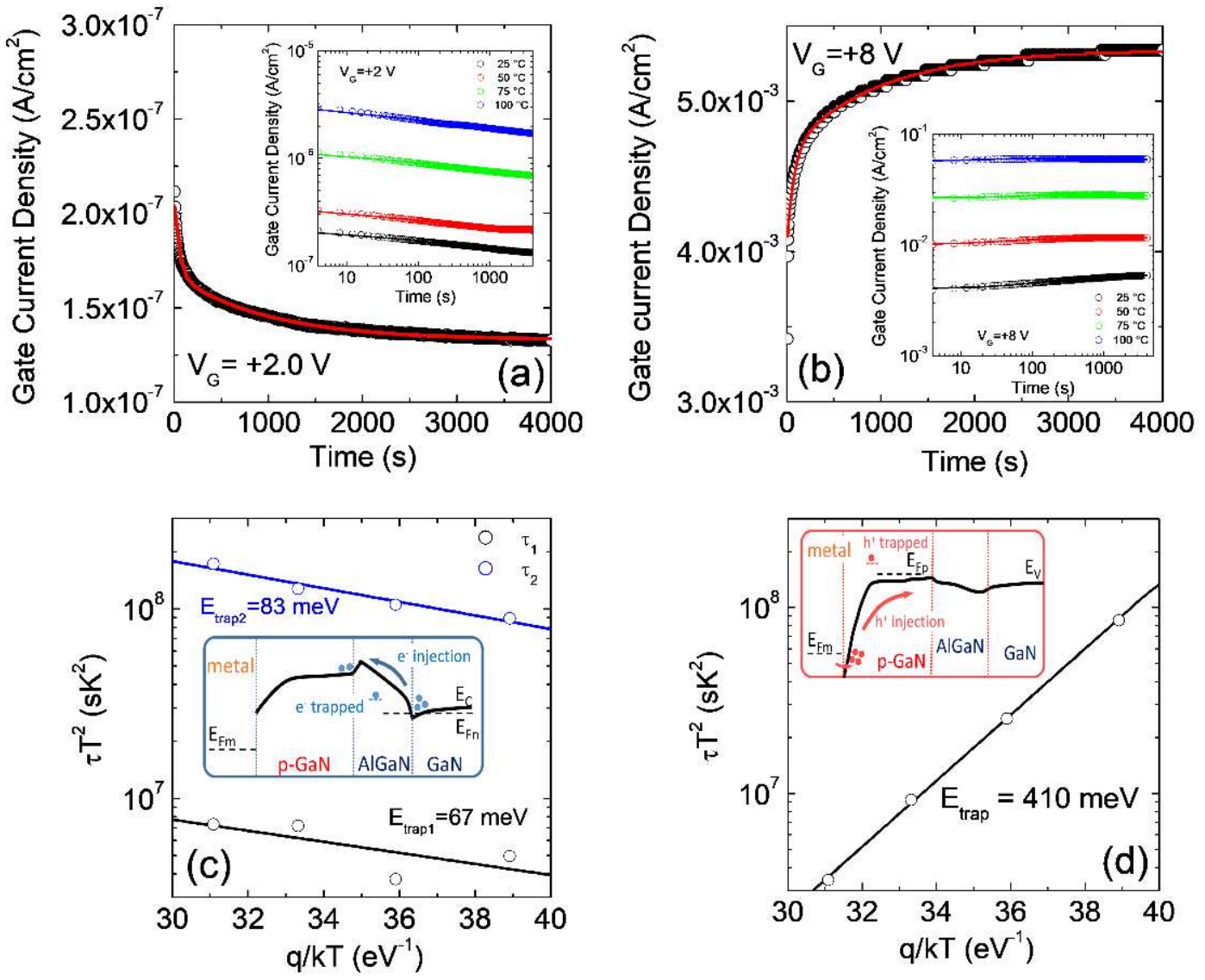

Fig. 3: Gate current density transient ($J_G$-t) collected at $V_G$ = +2 V (a) and $V_G$ = +8 V (b) with the fits obtained using Eq.1. The insets show the gate current density transient collected at different temperatures in the two cases. The relative Arrhenius plot of the time constants carried out using Eq.5 for $V_G$ = +2 V (c) and $V_G$ = +8 V (d). The insets show a schematic band diagram illustrating the trapping effects of electrons (c) or holes (d) during gate stress measurement.

Besides discussing on the origin of the $V_{TH}$ instability observed at low and high gate bias stress, it is also important to clarify the cause of the decrease of $J_G$ observed after long stress for $V_G$= +9.0 V (after 360 s) and $V_G$= +9.5 V (after 240 s). Indeed, at these bias levels, after a first increase, $J_G$ starts to decrease, thus meaning that the initial positive charge trapping is overtaken by a net negative charge trapping. The origin of this phenomenon can be identified with the generation of new traps for electrons. In order to shed light on the nature of these generated traps, a selected number of devices have been stressed at $V_G$= +9.5 V, by limiting the gate current value below $1\times10^{-3}$ A in order to avoid catastrophic breakdown event. Once the "soft-breakdown" event is induced, Emission Microscopy (EMMI) analyses were used to detect and to localize precisely the region where the event occurred. A similar approach has been recently used to assess the reliability of silicon carbide transistors [35]. Then, focussed ion beam (FIB) was used to prepare lamellas in the region of interest. Fig. 4(a) shows the presence of narrow defects in the cross section STEM image collected along the (11-20) direction, which can be regarded as "dislocation-like" defects since they resemble the edge dislocation features. Such dislocation-like defect differs from all the other native dislocations usually observed by STEM imaging. In fact, it is straight and very narrow. It starts from the metal/p-GaN interface, where the soft-breakdown event creates a partial detachment of the metal from the p-GaN surface (as can be seen in Fig. 4(a)), and crosses the AlGaN layer up to end at the underlying GaN layer. High Resolution - High Angle Annular Dark Field (HR-HAADF) STEM image of the dislocation-like defect, reported in Fig. 4(b), shows that the crystalline structure is perturbed in the defective region, and the width of the generated defect corresponds to about 3 atomic planes. The created strain effect is compatible either with the formation of partial extra-plane in the defect region or with the formation of atomic vacancies (both Gallium and Nitrogen) or interstitials. The high holes injection occurring at these bias level could be the origin of these localized defects, which act as electron-trapping states during device stress. To determine the activation energy of these traps, current gate transient at different temperature (between 25 °C and 100 °C) have been acquired at fixed stress bias of $V_G$ = +9 V, as reported in Fig. S1(a) of the Supplementary Information Hence, with a similar approach as that used for gate bias stress at +2V and +8V, it was possible to draw an Arrhenius plot and estimate an activation energy of 305 meV (see Fig. S1(b) of the Supplementary Information). Indeed, values of activation energy close to 300 meV have been often reported for extended defects in AlGaN/GaN heterostructure, like treading dislocations [36,37,38].

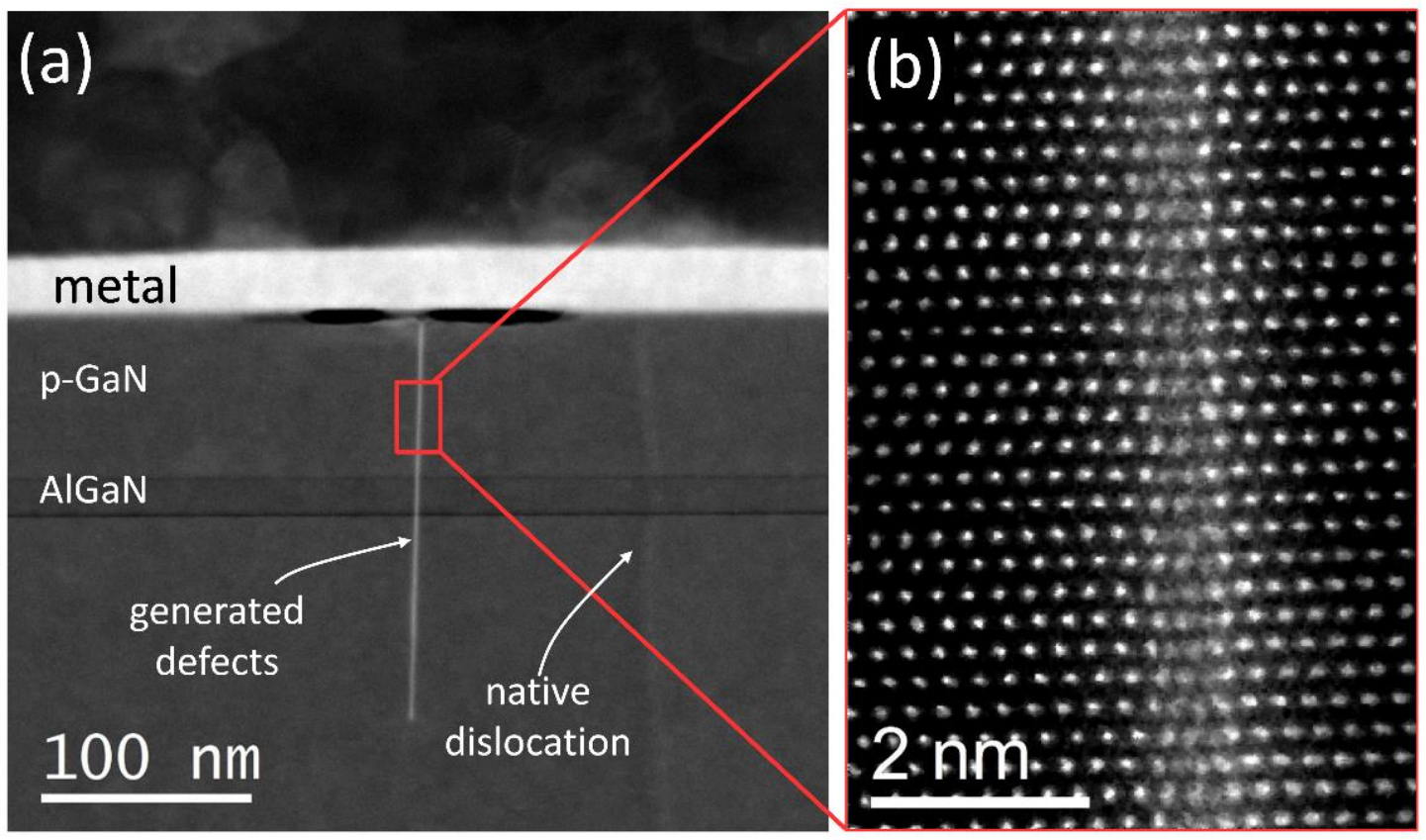

Fig. 4: Cross section STEM image of the region identified by EMMI analysis across the (11-20) direction (a), HRSTEM of the observed generated "dislocation-like" defect (b)

In conclusion, the $V_{TH}$ instability and gate current density transient have been monitored at different prolonged gate bias stress to shed light on the charge trapping phenomena occurring at the gate region of p-GaN HEMTs. The gate current transients could be correlated with electrons or holes trapping, depending on the gate bias stress level. In particular at low bias stress ($V_G$= 2-6 V) trapped electrons caused a decrease of the gate current, while at high stress bias ($V_G$= 7-9.5 V) trapped holes produces an increase of the gate current density. The time constants associated to the decrease or increase of the gate current have been estimated in case of low- and high-bias stress. The temperature dependence of the time constant was used to estimate the energy trap level in the case of $V_G$= 2 V and $V_G$= 8 V. In particular, electron trapping occurred at lower gate bias ($V_G$= 2 V) has been correlated with two trap levels close to the conduction band ($E_{trap1}$ =67 meV and $E_{trap2}$= 83meV). Instead, at $V_G$ = 8 V, a trap state of 400 meV above the valence band was estimated for the hole trapping. In addition, the generation of new traps for electrons causing a change in the $J_G$ behaviour has been observed at higher gate bias stress (8 V $<V_G<$ 9.5 V). The nature of this traps was explained by structural analysis that revealed the generation of a "dislocation-like" defect in the heterostructures below the metal gate.

#### Supplementary Material

See supplementary material, for the gate current transients analysis acquired in the p-GaN HEMTs at high bias stress (i.e. at $V_G$= +9 V). The relative Arrhenius plot of the time constants is also displayed for traps activation energy estimation.

#### Acknowledgements

This work has been carried out in the framework of the European Project GaN4AP (Gallium Nitride for Advanced Power Applications). The project has received funding from the Electronic Component Systems for European Leadership Joint Undertaking (ECSEL JU), under grant agreement No.101007310. This Joint Undertaking receives support from the European Union's Horizon 2020 research and innovation programme, and Italy, Germany, France, Poland, Czech Republic, Netherlands.

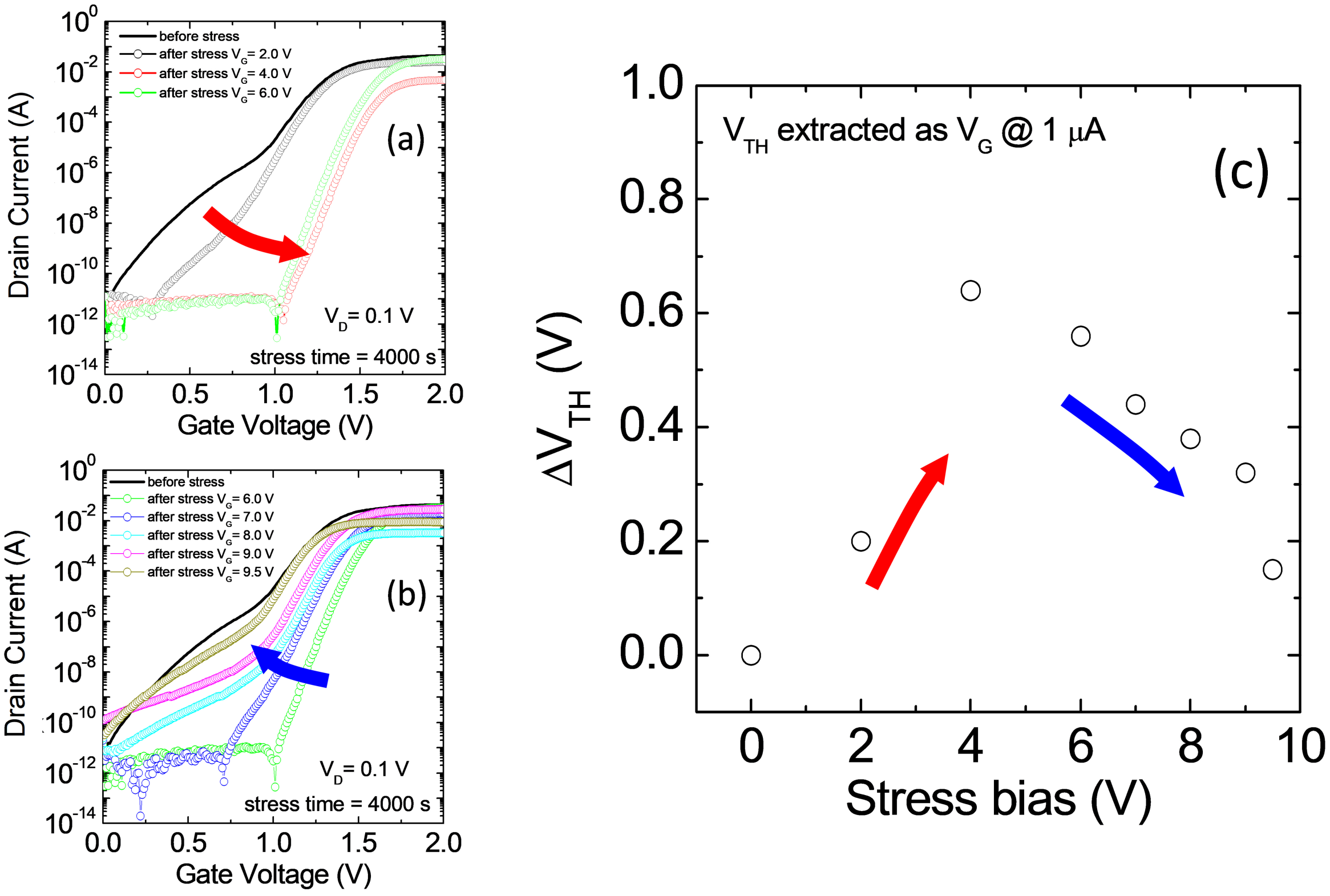
before stress
after stress V_G = 2.0 V
after stress V_G = 4.0 V
after stress V_G = 6.0 V
(a)
V_D = 0.1 V
stress time = 4000 s
Drain Current (A)
Gate Voltage (V)
before stress
after stress V_G = 6.0 V
after stress V_G = 7.0 V
after stress V_G = 8.0 V
after stress V_G = 9.0 V
after stress V_G = 9.5 V
(b)
V_D = 0.1 V
stress time = 4000 s
Drain Current (A)
Gate Voltage (V)
V_TH extracted as V_G @ 1 μA
(c)
ΔV_TH (V)
Stress bias (V)

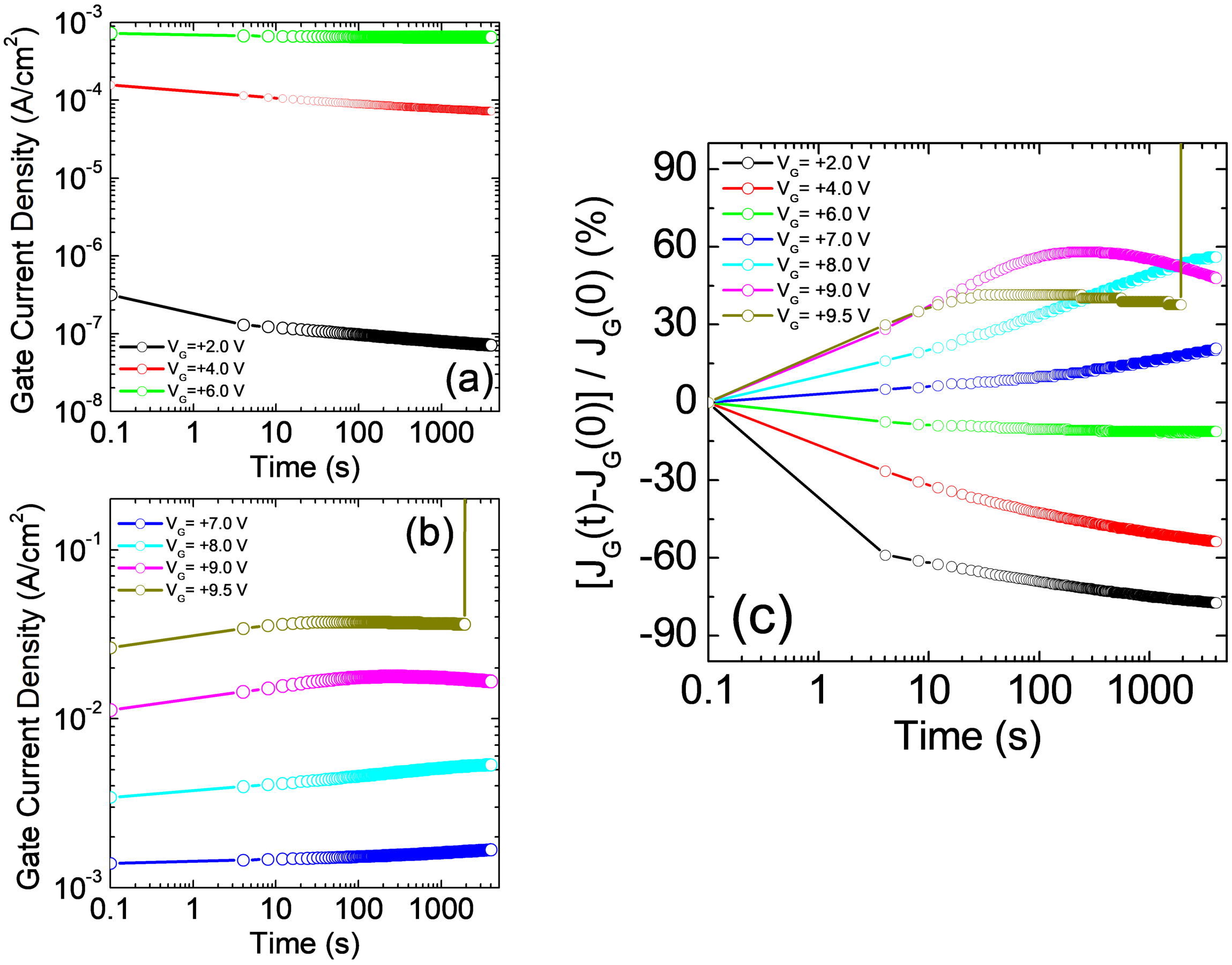

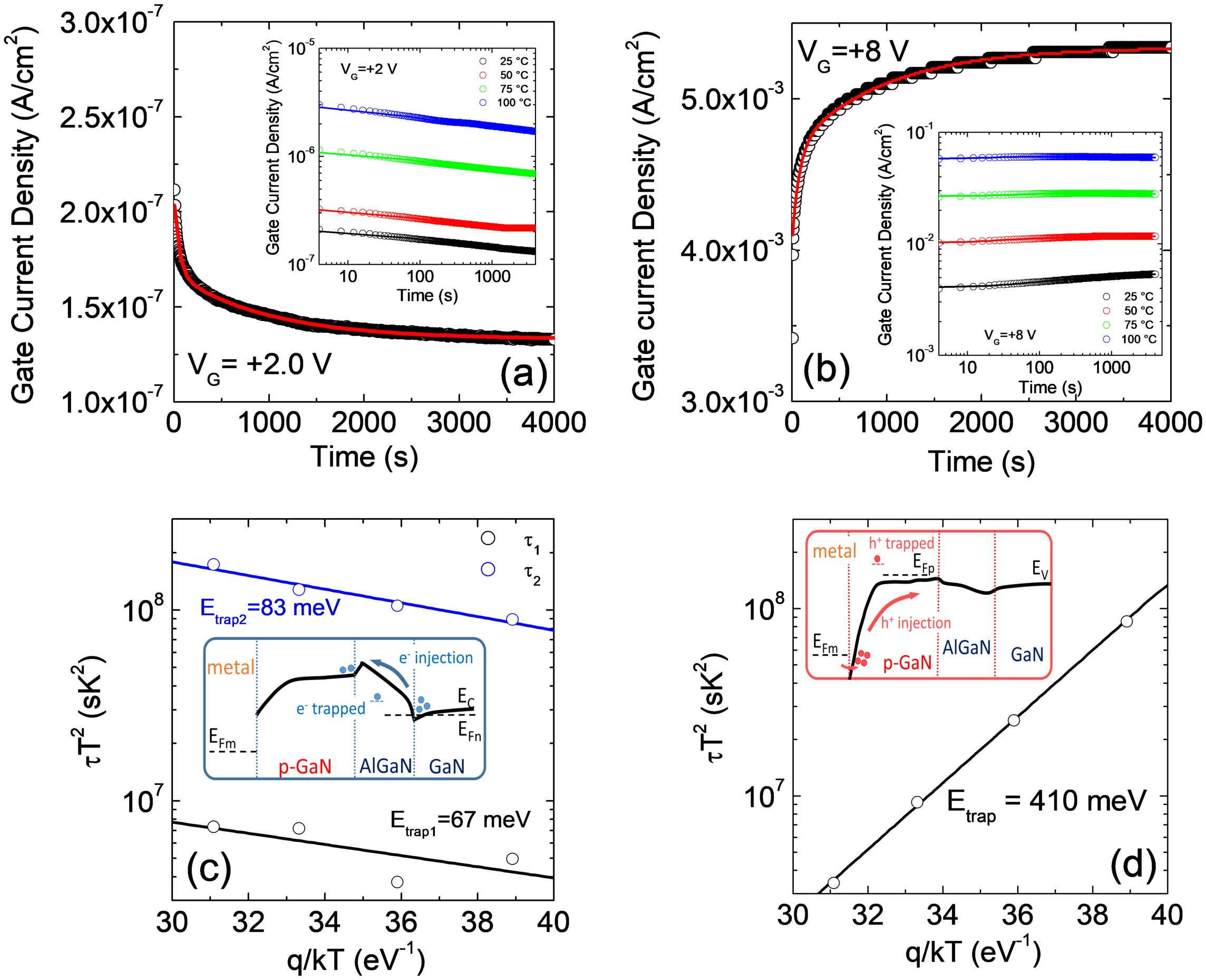

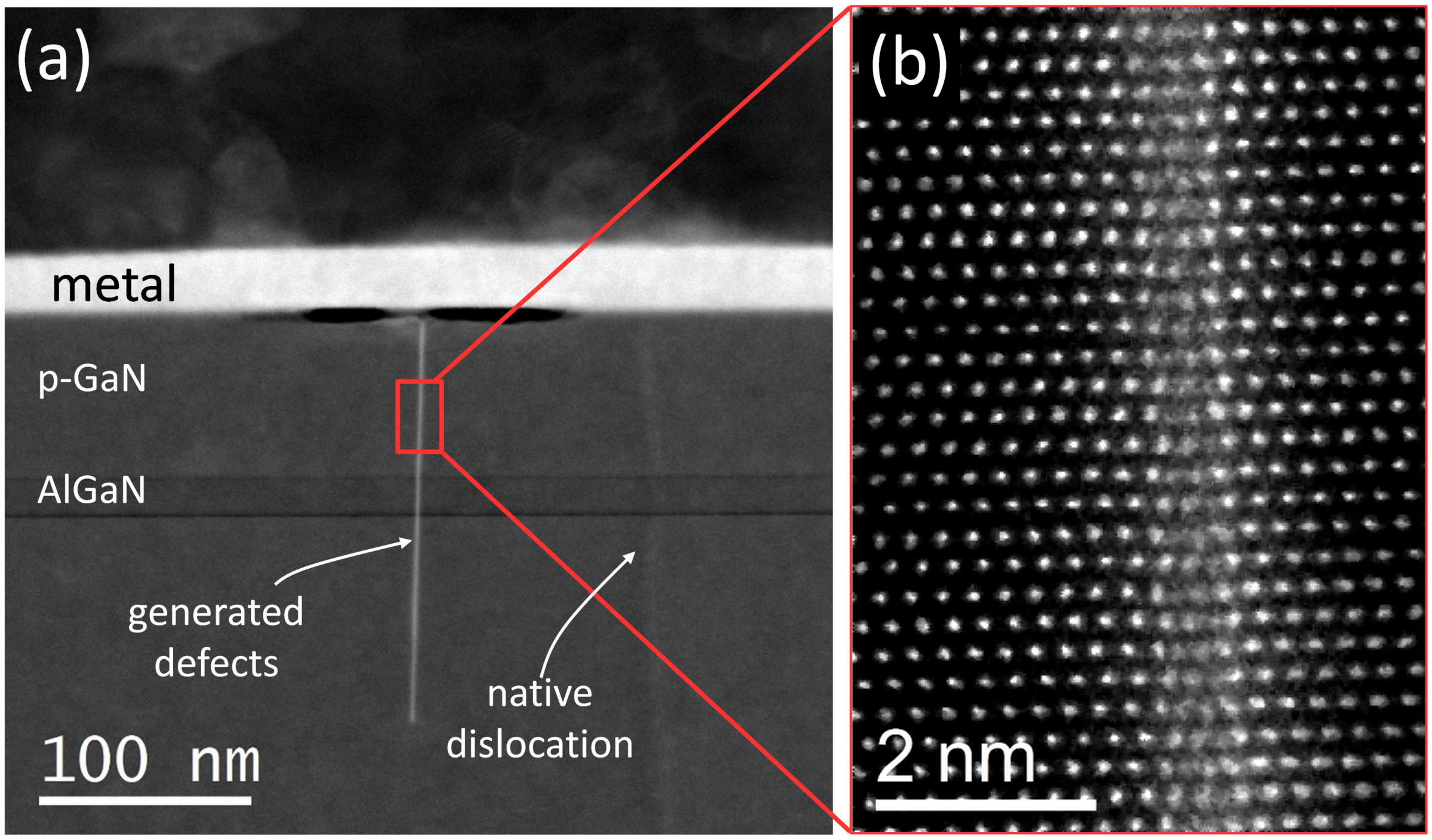